\documentclass[pre,showpacs, twocolumn,
               groupedaddress,floatfix,a4paper,10point]
              {revtex4}
\usepackage[utf8]{inputenc}
\usepackage[T1]{fontenc}
\usepackage{graphicx}
\usepackage{amssymb}
\usepackage{amsmath}

\usepackage{relsize}
\usepackage{color}
\usepackage{times}
\usepackage{array}
\usepackage[left=2.0cm,right=1.2cm,top=2cm, bottom=2cm]{geometry}

\usepackage{dsfont}

\newcommand{\avg}[1]{{\left<#1\right>}}

\begin{document}

\title{Trust transitivity in social networks}

\author{Oliver Richters}
\author{Tiago P. Peixoto}
\email[]{tiago@fkp.tu-darmstadt.de}
\affiliation{Institut für Festkörperphysik, TU Darmstadt, Hochschulstrasse 6,
  64289 Darmstadt, Germany}

\date{\today}

\begin{abstract}
  Non-centralized recommendation-based decision making is a central
  feature of several social and technological processes, such as market
  dynamics, peer-to-peer file-sharing and the web of trust of digital
  certification. We investigate the properties of trust propagation on
  networks, based on a simple metric of trust transitivity. We
  investigate analytically the percolation properties of trust
  transitivity in random networks with arbitrary degree distribution,
  and compare with numerical realizations. We find that the existence of
  a non-zero fraction of \emph{absolute trust} (i.e. entirely confident
  trust) is a requirement for the viability of global trust propagation
  in large systems: The average pair-wise trust is marked by a
  discontinuous transition at a specific fraction of absolute trust,
  below which it vanishes. Furthermore, we perform an extensive analysis
  of the Pretty Good Privacy (PGP) web of trust, in view of the concepts
  introduced. We compare different scenarios of trust distribution:
  community- and authority-centered. We find that these scenarios lead
  to sharply different patterns of trust propagation, due to the
  segregation of authority hubs and densely-connected communities. While
  the authority-centered scenario is more efficient, and leads to higher
  average trust values, it favours weakly-connected ``fringe'' nodes,
  which are directly trusted by authorities. The community-centered
  scheme, on the other hand, favours nodes with intermediate degrees, in
  detriment of the authorities and its ``fringe'' peers.
\end{abstract}


\maketitle

\section{Introduction}

Several social and technological systems rely on the notion of trust, or
recommendation, where agents must make their decision based on the
trustworthiness of other agents, with which they interact. One example
are buyers in markets~\cite{nicholaas_j._vriend_self-orgainzed_1994,
  bornholdt_handbook_2003}, who may share among themselves their
experiences with different sellers, or lenders which may share a belief
that a given borrower will not be able to pay
back~\cite{anand_financial_2009}.  Another example are peer-to-peer
file-sharing programs~\cite{kamvar_eigentrust_2003,
  bornholdt_handbook_2003}, which often must know, without relying on a
central authority, which other programs act in a fair manner, and which
act selfishly. In the same line, an even more direct example is the web
of trust of digital certification, such as the Pretty Good Privacy (PGP)
system~\cite{guardiola_macro-_2002, boguna_models_2004}, where regular
individuals must certify the authenticity of other individuals with
digital signatures. In all these systems, the agents lack global
information, and must infer the reliability of other agents, based
solely on the opinion of trusted peers, thus forming a network of trust.
In this paper, we present an analysis of trust propagation based on the
notion of \emph{transitivity}: If agent $a$ trusts agent $b$, and agent
$b$ trusts agent $c$, then, to some extent, agent $a$ will also trust
agent $c$. Based on this simple concept, we define a trust metric with
which the reliability of any reachable agent may be inferred. Instead of
concentrating on the minutiae of trust propagation semantics, we focus
on the topological aspect of trust networks, using concepts from network
theory~\cite{newman_networks:_2010}. Using random networks as a simple
model, we investigate the necessary conditions for trust to
``percolate'' through an entire system. We then apply the concepts
introduced to investigate in detail the PGP web of trust, possibly the
best ``real'' example of a trust propagation system, which is completely
accessible for investigation. We focus on the role of the strongly
connected nodes in the network --- the so called \emph{trust
  authorities} --- which represent a different paradigm of trust
delegation, in comparison to the decentralized community-based approach,
which is also heavily present in the network.

This paper is divided as follows. In section~\ref{sec:metric} we define
the trust metric used; in section~\ref{sec:percolation} we consider the
problem of trust percolation in random networks with different trust
weight distributions. In in section~\ref{sec:pgp} we turn to the
analysis of the PGP network, and provide an extensive analysis of
topology of the PGP network, and of trust propagation according to
different trust distribution scenarios. Finally, in
section~\ref{sec:conclusion} we provide some final remarks and a
conclusion.

\section{Trust metric}\label{sec:metric}

Trust is the measure of belief that a given entity will act as one
expects. It is often associated with positive, desirable attributes, but
it may not always be the case (e.g. one may have trust that someone will
act undesirably). Humans use trust to make decisions when more direct
information is unavailable. In general, humans will decide their level
of trust based on arbitrary, heuristic rules, since there is no formal
consensus on how to evaluate trust. We will deliberately avoid the
detailed formalization of these rules, and instead rely on two
simplifications: 1. We will treat trust simply as a probability that a
given assessment about an agent is true or false (e.g. fair/reliable or
not); 2. We further assume that this belief is \emph{transitive},
i.e. if agent $a$ trust agent $b$, which in turn trusts agent $c$, then
$a$ will also trust $c$, to some extent. This makes trust propagation
easier to analyse, while retaining the most intuitive properties of
trust propagation.

\begin{figure}[hbt]
  \includegraphics*{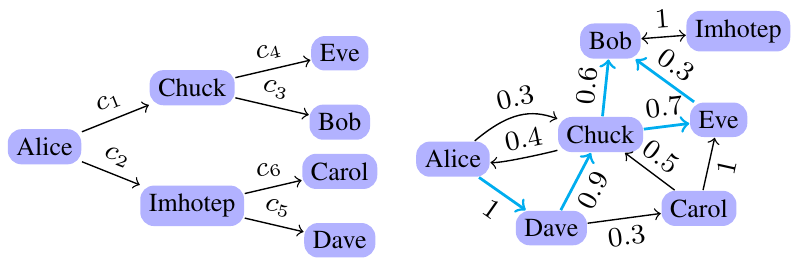}
\caption{Examples of trust networks: {\bf Left:} A directed tree. {\bf
    Right:} A more realistic example. The edges in blue are the ones
  which contribute to the value of trust from Bob to Alice, according to
  Eq.~\ref{eq:pervasive}.\label{fig:net_ex}}
\end{figure}

\begin{figure}[hbt]
  \includegraphics*{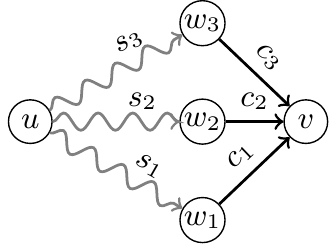}
\caption{Illustration of the paths used to calculate $t_{u,v}$ according to
  Eq.~\ref{eq:pervasive}. The vertices $w_i$ are the in-neighbours of $v$, and
  the values $s_i = s^{G\setminus \{v\}}_{u,w_i}$ are the values of best trust
  (Eq.~\ref{eq:lotta}) from $u$ to $w_i$, with vertex $v$ removed from the
  graph.\label{fig:pervasive}}
\end{figure}

We will consider a system of $N$ agents which form a directed trust
network: Each agent $v$ (represented by a vertex, or node) has a number
of interactions (represented by directed edges, or links) with other
agents $\{u_i\}$ for which a value $c_{v,u_i} \in [0,1]$ of \emph{direct
  trust} is defined {\it a priori}, and which can be interpreted as a
probability. This value represents a direct experience agent $v$ had
with $u_i$, which is not inferred from any other agent. We then define
the \emph{inferred trust} $t_{ij} \in [0,1]$ from agent $i$ to any agent
$j$, which is somehow based on the values of $c_{v,u_i}$. In a simple
situation where there is only one possible path between any two given
nodes (i.e. the network is a directed tree, as the example on the left
in Fig.~\ref{fig:net_ex}), one could simply multiply the values of $c$
along the single path to obtain $t$, e.g. $t_{\text{Alice},\text{Bob}} =
c_1c_3$, in the example of Fig.~\ref{fig:net_ex}. In general, however,
the situation may be more complicated, as in the example on the right of
Fig.~\ref{fig:net_ex}, where there is a variety of possible (often
``contradictory'') transitive paths between most pairs of nodes.
Perhaps the simplest way of defining a trust metric would be to consider
only the \emph{best} transitivity path between two nodes, i.e., the one
where the trust transitivity is maximum,
\begin{equation}
  s_{u,v} = \max\left\{\prod_{\{e_i\}} c_{e_i} \right\},
  \qquad \forall\,\{e_i\} \in P_{u\rightsquigarrow v},
  \label{eq:lotta}
\end{equation}
where $P_{u\rightsquigarrow v}$ is the set of all paths from $u$ to $v$,
$\{e_i\}$ is the set of edges in a given path, and $c_e$ is the direct
trust associated with a given edge. This definition is an attractive
one, since it corresponds directly to the concept of minimum distance on
weighted graphs, which is defined as the sum of weights along the path
with the smallest sum. This is easily seen by noticing that
$\prod_{\{e_i\}} c_{e_i} = \exp\{\sum_{\{e_i\}} \omega_{e_i}\}$, with
$\omega_{e_i}=-\ln c_{e_i} \ge 0$ being the edge weights (with the special
value of $\omega_{e_i} = \infty$ if $c_{e_i}=0$). However, it is clear that
this approach leads to an optimistic bias, since the best path obviously
favors large values of trust, and uses only a small portion of the
information available in the network. As an illustration consider the
network on the right of Fig.~\ref{fig:net_ex}, where the value of
$s_{\text{Alice},\text{Bob}}$ is $1\times0.9\times0.6 = 0.54$, via Dave
and Chuck. However, if Chuck is directly consulted, the transitivity
drops to $0.3\times0.6 = 0.18$. In principle, there is no reason to
prefer any of the two assessments over the other. One may attempt to
rectify this by considering instead \emph{all} possible paths between
two nodes,
\begin{equation}
  \tilde{t}_{u,v} = \frac{\displaystyle\sum_{u\rightsquigarrow v}\omega_{u\rightsquigarrow v}\prod_{e \in u\rightsquigarrow v} c_e}
                         {\displaystyle\sum_{u\rightsquigarrow v}\omega_{u\rightsquigarrow v}},
 \label{eq:rome}
\end{equation}
where $\omega_{u\rightsquigarrow v}$ is a weight associated with a given
path $u\rightsquigarrow v$. It should be chosen to minimize the effect
of a very large number of paths with very low values of trust, without
introducing an optimistic bias on the final trust value. One apparently
good choice is to consider the transitivity value of the path itself,
but not including the last edge,
\begin{equation}\label{eq:weights}
 \omega_{u\rightsquigarrow v} = \prod_{e \in  u\rightsquigarrow v} c_e + (1-c_e)\delta(e,e_{\to v}),
\end{equation}
where $e_{\to v}$ is the last edge in the path, and $\delta$ is the
Kronecker delta. Not only this avoids a bias in the final value of
$\tilde{t}_{u,v}$, but also $\omega_{u\rightsquigarrow v}$ has a simple
interpretation as being the value of trust on the \emph{final}
recommendation, which is completed by the last edge. While this may seem
reasonable, and uses all available information in the network, it has
two major drawbacks: 1. It is very computationally costly to consider
all possible paths between two nodes, even in moderately sized
networks. It would represent an unreasonable effort on part of the
agents to use all this information. 2. Computed as in Eq.~\ref{eq:rome},
the value of $\tilde{t}_{u,v}$ has the unsettling behaviour of tending
to zero, whenever the number of paths become large (as they often are),
even when paths are differently weighted. Consider a simple scenario
where the network is a complete graph, i.e. all possible edges in the
network exist, and all of them have the same direct trust value
$c$. Since there are ${N-2 \choose l} l!$ paths of length $l+1$ between
any two vertices, the value of inferred trust between any two nodes can
be calculated as
\begin{align}
  \tilde{t}_{u,v} &= \frac{\sum_{l=0}^{N-2}{N-2 \choose l} l! c^{2l+1}}
                        {\sum_{l=0}^{N-2}{N-2 \choose l} l! c^{l}} \\
                 &\leq c^{N-1} \exp\left(\frac{1}{c^2} - \frac{1}{c}\right),
\end{align}
from which it is easy to see that $\lim_{N\to\infty} \tilde{t}_{u,v} =
0$ for $c < 1$. This is an undesired behavior, since one would wish that
such highly connected topologies (which often occur as subgraphs of
social networks, known as \emph{cliques}) would result in \emph{higher}
values of trust. In order to compensate for this one would have to use a
more aggressive weighting of the possible paths. We propose the
following modification, which combines some features of both previous
approaches: Instead of considering all possible paths, we consider only
those with the largest weights to all the in-neighbours of the target
vertex, as shown in Fig.~\ref{fig:pervasive}. This leads to a trust
metric defined as
\begin{equation}\label{eq:pervasive}
  t_{u,v} = \frac{\sum_w A_{w,v} \left(s^{G\setminus \{v\}}_{u,w}\right)^2c_{w,v}}
  {\sum_w A_{w,v} s^{G\setminus \{v\}}_{u,w}},
\end{equation}
where the path weights are the best trust transitivity to the
in-neighbours, $s^{G\setminus \{v\}}_{u,w}$, which are calculated after
removing the target vertex from the graph (so that it cannot influence
its own trust).
We call this trust metric \emph{pervasive trust}, and it corresponds to
the intuitive strategy of searching for the nodes with a direct
interaction with the target node (the final arbitrators), and weighting
their opinions according to the best possible trust transitivity leading
to them. It can be seen that this definition does not suffer form the
same problems of Eq.~\ref{eq:rome}, again by considering the same
complete graph example, with uniform direct trust $c$. Since in this
situation every target vertex has $N-2$ in-neighbours different from the
source, and the shortest path to each of these in-neighbours is of
length one, the value of pervasive trust can be easily calculated as
\begin{equation}
  t_{u,v} = \frac{(N-2) c^3 + c}{(N-2) c + 1},
\end{equation}
which converges to $t_{u,v} \approx c^2$ for $N \gg 1$. Thus the indirect
opinions with value $c^2$ dominate the direct trust value $c$, but the inferred
value does not vanish, as with the definition of Eq.~\ref{eq:rome}. Considering
again the example on the right of Fig.~\ref{fig:net_ex}, we obtain the value
$t_{\text{Alice,Bob}} = (0.9^2 \times 0.6 + (0.9\times0.7)^2 \times 0.3) / (0.9
+ 0.9\times0.7) \approx 0.4$, from the edges outlined in blue in the
figure. Additionally, the definition of pervasive trust works as one would
expect in the trivial example on the left of Fig.~\ref{fig:net_ex}, where
$s_{u,v}$ and $t_{u,v}$ have the same values.

We note that the numerical computation of $s_{u,v}$ can be done by using
Dijkstra's shortest path algorithm~\cite{dijkstra_note_1959,
  brandes_network_2005}, which has a complexity of $O(N\log N)$. Thus
the entire matrix $s_{u,v}$ can be calculated in $O(N^2\log N)$
time. The same algorithm can be used to calculate $t_{u,v}$, but since
each target vertex needs to be removed from the graph, and thus a new
search needs to be made for each different target, this results in
$O(N^3\log N)$ time. It is possible to improve this by performing
searches in the \emph{reversed} graph, i.e., for each target vertex $v$,
the contribution to $t_{u,v}$ from all sources $u$ can be calculated
simultaneously, after $v$ is removed, by performing a single reversed
search from each of the in-neighbours of $v$ to each source $u$. This
way, the entire $t_{u,v}$ matrix can be computed in $O(kN^2\log N)$ time
(where $k=E/N$ is the average degree of the network), which is
comparable to the computation time of $s_{u,v}$ for sparse graphs.

\subsection{Comparison with other trust metrics}

Other trust metrics have been proposed in the literature, mainly by
computer scientists, seeking to formalize the notion of trust in
peer-to-peer computer systems. Some are quite detailed, like the usage
of subjective logic by Jøsang et al~\cite{jsang_trust_2006}, and others
are comparable with the simplistic approach taken in this work, such as
Eigentrust~\cite{kamvar_eigentrust_2003} and more recently
TrustWebRank~\cite{walter_personalised_2009}. These last metrics are
based on the notion of \emph{feedback
  centrality}~\cite{brandes_network_2005}, which are calculated by
solving some linear system. The Eigentrust metric requires the trust
network to be a stochastic matrix (i.e. the sum of the trust values of
the out-edges of all vertices must sum to unity) and the inferred trust
values are given by the steady state distribution of the corresponding
Markov chain (i.e. the left eigenvector of the stochastic matrix with
unity eigenvalue, hence the name of the metric). Thus the inferred trust
values are \emph{global} properties, independent of any source vertex
(i.e. non-personalized), which is non-intuitive. Additionally, the
requirement that the trust network is stochastic means that only
\emph{relative} values of trust are measured, and the absolute
information is lost. Furthermore, such an approach is strongly affected
by the presence of loops in the network, which get counted multiple
times, which is also non-intuitive as far as trust transitivity is
concerned. The metric TrustWebRank~\cite{walter_personalised_2009} tries
to fix some of these problems by borrowing ideas from the
PageRank~\cite{page_pagerank_1999} algorithm, resulting in a metric
which also requires a stochastic matrix, but is personalised. However,
in order for the algorithm to converge, it depends on the introduction
of an \emph{damping factor} which eliminates the contribution of longer
paths in the network, independently of its trust value. This is an {\it
  a priori} assumption that these paths are not relevant, and may not
correspond to reality. Additionally, the strange role of loops in the
network is the same as in the Eigentrust metric. However, since there is
no consensus on how a trust propagates, and the notion of trust lacks a
formal, universally accepted definition, in the end there is no
``correct'' or ``wrong'' metric. We only emphasize that our approach is
derived directly from the simple notion of trust transitivity, is easy
to interpret, propagates \emph{absolute} values of trust, and makes no
assumption whatsoever about the network topology, and direct trust
distribution.

\section{Trust percolation}\label{sec:percolation}

Trust transitivity is based on the multiplication of direct trust
values, which may tend to be low if the paths become long. Therefore, it
is a central problem to determine if the trust transitivity between two
randomly chosen vertices of a large network vanishes if the system
becomes very large. This provides important information about the
viability of trust transitivity on large systems. As a simple network
model, we will consider random directed networks with arbitrary degree
distributions~\cite{newman_random_2001}. We will also suppose that the
direct trust values in the range between $c$ and $c+dc$ will be
independently distributed with probability $\rho_c(c)dc$, where
$\rho_c(c)$ is an arbitrary distribution. The objective of this section
is to calculate the average best trust transitivity $\left<s\right>$,
given by Eq.~\ref{eq:lotta}, and the average pervasive trust
$\left<t\right>$, Eq.~\ref{eq:pervasive}, between randomly chosen pairs
of source and target vertices. In random networks, the value of average
pervasive trust will be given simply as $\left<t\right> = \left<s\right>
\left<c\right>$, since the best paths to the in-neighbours of a given
vertex are uncorrelated, and the probability that they pass through the
node itself tend to zero, in the limit of large network size. Therefore
we need only to concern ourselves with the average best trust
transitivity $\left<s\right>$.

Networks are composed of components of different types and sizes: For
each vertex there will be an \emph{out-component}, which is the set of
vertices reachable from it, and an \emph{in-component}, which is the set
of vertices for which it is reachable. A maximal set of vertices which
are mutually reachable is called a \emph{strongly connected
  component}. Random graphs often display a phase transition in the size
and number of these components: If the number of edges is large enough,
there will be the sudden formation of a giant (in-, out-, strongly
connected) component, which spans a non-vanishing fraction of the
network~\cite{newman_networks:_2010,newman_random_2001}. The existence
of these giant components is obviously necessary for a non-vanishing
value of trust to exist between most vertices, but it is not sufficient,
since it is still necessary that the multiplication of direct trust
values along most shortest paths do not become vanishingly small. As an
illustration, consider a sparse graph (i.e. with finite average degree),
with an arbitrary degree distribution. In the situation where there is a
sufficiently large giant out-component in the graph, the average
shortest path from a randomly chosen root vertex to the rest of the
network is given approximately~\cite{newman_random_2001} by
\begin{equation}\label{eq:pl}
  l \approx \frac{\ln(N/\left<k\right>)}{\ln(\left<k_2\right>/\left<k\right>)},
\end{equation}
independently of the out-degree distribution (as long as
$\left<k\right>$ and $\left<k_2\right>$ are finite positive), where $N$
is the number of vertices, $\left<k\right>$ is the average out-degree
and $\left<k_2\right>$ is the average number of second out-neighbours,
and it is assumed that $N \gg \left<k\right>$ and $\left<k_2\right> \gg
\left<k\right>$~\footnote{%
  An analogous expression for the distance from the entire network to a
  randomly chosen \emph{target} can be obtained by replacing
  $\left<k\right>$ and $\left<k_2\right>$ with the average in-degree and
  second in-neighbours, $\left<j\right>$ and $\left<j_2\right>$
  respectively.%
}. Since the edges are weighted, the average length of the best paths
can differ from $l$, but can never be smaller. Thus, an upper bound on
the average best trust is given by $\left<s\right> =
o(\max{\{c_i\}}^l)$, where $\max{\{c_i\}}$ is the maximum value of
direct trust in the network. In the situation where $\max{\{c_i\}} < 1$,
we have that $\lim_{N\to\infty} \left<s\right> = o(0)$, since
$\lim_{N\to\infty}l = \infty$. Therefore, if there are no values of
$c=1$ in the network, the average trust will always be zero in sparse
networks. The only possible strategies for non-vanishing values of
average trust is either to have a non-zero fraction of $c=1$ (which we
will call \emph{absolute trust}), or for the network to be dense, such
that $l$ remains finite for $N\to\infty$.

With the above consideration in mind, we now move to calculate the
average trust transitivity values. For that we modify the generating
function method used in~\cite{newman_random_2001} to obtain the
distribution of component sizes. The objective is to obtain a
self-consistency condition for the distribution of best trust
transitivity values by describing the direct neighbourhood of a single
vertex, which is based on the following observation: A randomly chosen
vertex $u$ with out-neighbours $w_i$, each with direct trust from $u$
given by $c_i$, will trust another randomly chosen vertex $v$ with a
best trust transitivity value of $s^+$ only if $\max(c_is_i^+) = s^+$,
where $s_i^+$ is the best trust transitivity from $w_i$ to $v$. In a
random network which is sufficiently large, $s^+$ and $s_i^+$ should
both be drawn from the same distribution $\rho^+_s(s)$. This gives a
self-consistency condition for $\rho^+_s(s)$ which is given
schematically as follows, \resizebox{\columnwidth}{!}{
  \includegraphics*{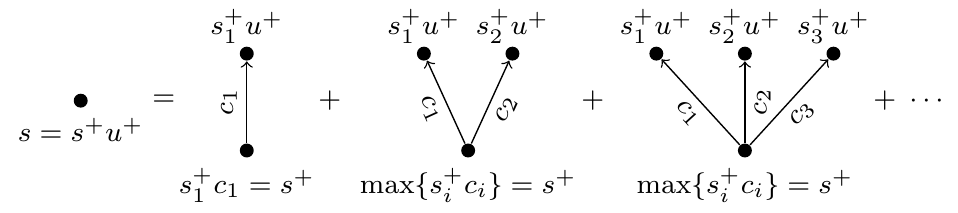}} where each term
corresponds to the probability of the vertex having a given number of
out-neighbours, and the maximum best trust transitivity being equal the
desired value. Note that we have explicitly multiplied every instance of
$s^+$ with a arbitrary free variable $u^+$, which cannot be determined
by the above self-consistency alone, and has to be described
separately. Each term on the right is weighted by the out-degree
probability $p_k$. In terms of the cumulative distribution
$\tilde{\rho}^+_s(s) = \int_0^sdu \rho^+_s(u)$, this self-consistency
can be expressed as
\begin{equation} \label{eq:self-cons+}
    \tilde{\rho}^+_s (s) = \sum_k p_k[\tilde{\beta}^+(s)]^k,
\end{equation}
where $\tilde{\beta}^+(x)$ is the cumulative probability that $s^+c <
x$, with $c$ distributed by $\rho_c(c)$, given by
\begin{equation}
    \tilde{\beta}^+ (s) = \int_0^1 dx \rho_c(x)\tilde{\rho}^+_s(s/x).
\end{equation}
The distribution $\rho^+_s(s)$ above does not equal $\rho_s(s)$, due to
the remaining variable $u^+$, for which one must still find an
appropriate distribution. This last piece is obtained by realizing that
$\rho_s(s)$ must also be subject to a complementary self-consistency
condition in the \emph{opposite} direction, following the in-neighbours:
A randomly chosen vertex $u$ with in-neighbours $w_i$, each with direct
trust to $u$ given by $c_i$, will \emph{be trusted} by another randomly
chosen vertex $u$ with a best trust value of $s^-$ only if
$\max(c_is_i^-) = s^-$, where $s_i^-$ is the best trust from $w_i$ to
$u$. This results in an entirely analogous self-consistency condition
for $\rho_s^-(s)$, where the out-degree distribution $p_k$ is replaced
by the in-degree distribution $p_j$. Since this last self-consistency is
also complete up to a free variable $u^-$, we can formulate the ansatz
that $u^+ = s^-$ and $u^- = s^+$, such that
\begin{equation}
  s = s^+ s^-.
\end{equation}
With this connection it is possible to obtain $\rho_s(s)$ as
\begin{align}
  \rho_s(s) =& \int_0^1du\rho^+_s(u)\rho^-_s(s/u) / u, \;\;\text{or} \\
            =& \int_0^1du\rho^+_s(s/u)\rho^-_s(u) / u,
\end{align}
and the average $\left<s\right>$ more directly as
\begin{align}
  \left<s\right> =& \int_0^1\int_0^1 ds^- ds^+ s^-s^+  \rho^-_s(s^-)\rho^+_s(s^+)\\
                 =& \left<s^-\right> \left<s^+\right>. \label{eq:avg}
\end{align}

By rewriting Eq.~\ref{eq:self-cons+} in terms of the generating functions
of the in- and out-degree distributions,
\begin{equation}
  G(z) = \sum_j p_j z^j \qquad F(z) = \sum_k p_k z^k,
\end{equation}
one obtains the self-consistency equations in a more compact form,
\begin{align}
    \tilde{\rho}^-_s (s) &= F(\tilde{\beta}^-(s)) \label{eq:self-cons-gf-} \\
    \tilde{\rho}^+_s (s) &= G(\tilde{\beta}^+(s)). \label{eq:self-cons-gf+}
\end{align}
These are integral equations, for which there are probably no general
closed form solutions. However, it is possible to solve them numerically
by successive iterations from an initial distribution, which we chose as
$\tilde{\rho}^0(s) = \Theta(s-1)$, where $\Theta(x)$ is the Heaviside
step function. From the numerical solutions the average values can be
obtained as $\left<s^-\right> = \int_0^1ds\rho^-_s (s)s =
1-\int_0^1ds\tilde{\rho}^-_s (s)$ (where the last expression is obtained
by integration by parts), and in analogous fashion for
$\left<s^+\right>$. The average value of best trust transitivity
$\left<s\right>$ is then given by Eq.~\ref{eq:avg}.

We turn now to the conditions necessary for non-vanishing average trust
transitivity.  Both Eqs.~\ref{eq:self-cons-gf-} and
\ref{eq:self-cons-gf+} accept the trivial solution
$\tilde{\rho}^{-/+}_s(s) = \Theta(s)$, which corresponds to
$\rho^{-/+}_s(s) = \delta(s)$, i.e. the average best trust is zero. As
discussed previously, for other solutions to be possible, we need to
consider a non-vanishing fraction of edges with absolute trust $c=1$ in
the network. Here we will consider direct trust distributions of the
form,
\begin{equation}
  \rho_c(c) = \gamma\delta(c-1) + (1-\gamma)\rho'_c(c),
\end{equation}
which correspond to a fraction $\gamma$ of edges with $c=1$, and a
complementary fraction $(1-\gamma)$ with $c$ given with probability
density $\rho'_c(c)$. We will consider two different versions of
$\rho'_c(c)$: A uniform distribution $\rho'_c(c) = 1$, and a
single-valued distribution $\rho'_c(c) = \delta(c-\eta)$, with $\eta =
1/2$. We will use two different degree distributions, the Poisson and
Zipf~\footnote{Many real networks have broad degree distributions with a
  power-law tail, $p_k \sim k^{-\tau}$ for large $k$. These networks are
  called \emph{scale-free}. The PGP network considered below is also an
  example of this.  }, and their respective generating functions,
\begin{align}
  p_j = \frac{\left<j\right>^j e^{-\left<j\right>}}{j!} &\qquad G(z) = e^{\left<j\right>(z-1)} \\
  p_j = \frac{j^{-\tau}}{\zeta(\tau)}  &\qquad G(z) = \frac{\text{Li}_{\tau}(z)}{\zeta(\tau)},
\end{align}
where $\zeta(\tau)$ is the Riemann $\zeta$ function, and
$\text{Li}_{n}(x)$ is the $n$th polylogarithm of $x$. For simplicity, we
will consider only the situation where $p_j = p_k$, and both $j$ and $k$
are independently distributed.
\begin{figure}[hbt]
  \includegraphics*{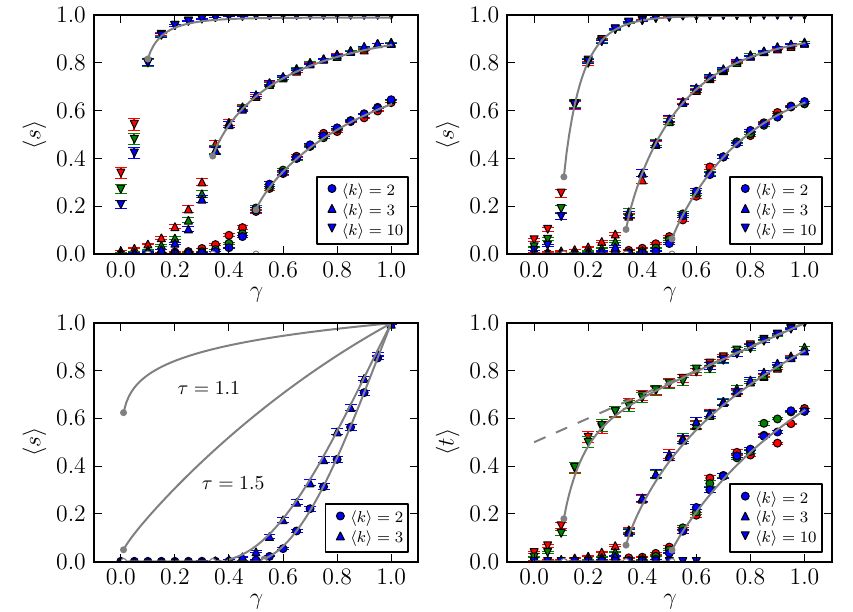}
  \caption{\label{fig:perc} Average values of best trust
    $\left<s\right>$ and pervasive trust $\left<t\right>$ as a function
    of the fraction of edges with absolute trust $\gamma$. {\bf Top
      left:} Networks with Poisson in- and out-degree distributions, and
    uniform trust distribution. {\bf Top right and bottom right:}
    Poisson distribution, and single-valued trust distribution. {\bf
      Bottom left:} Zipf distribution, and single-valued trust
    distribution. Solid lines correspond to analytical solutions, and
    symbols to numerical realizations of several networks of different
    sizes: $10^4$ (red), $10^5$ (green) and $10^6$ (blue) nodes. The
    dashed line shows the average direct trust $\left<c\right> = (\gamma
    + 1) / 2$.}
\end{figure}
In Fig.~\ref{fig:perc} are plotted the values of $\left<s\right>$ and
$\left<t\right>$, as a function of $\gamma$, for the different
distributions. It is also compared with numerical computations on actual
network realizations of different sizes. The main feature observed is a
first-order transition from vanishing trust to positive trust, at
specific values of $\gamma$. The transition values $\gamma^*$ correspond
exactly to the critical values of the formation of a giant component of
the induced subgraph composed only of edges with $c=1$, which has
average degree $\gamma\left<k\right>$~\cite{newman_random_2001}. For
graphs with Poisson degree distribution, this corresponds to
$\gamma^*=1/\left<k\right>$. It is worth observing that on finite
graphs, the average trust does not vanish very rapidly, and is still
non-zero for relatively large networks with $N=10^6$ nodes, even when
$\gamma=0$. This is attributed to the so-called small-world effect where
the average shortest path scales slowly as $l\sim\ln N$, as in
Eq.~\ref{eq:pl}. Therefore in practical situations where networks are
large but finite, $\gamma > \gamma^*$ it is not a strictly necessary
condition for system-wide trust propagation. Another interesting feature
is the behaviour of the average trust in graphs with Zipf degree
distribution. There, the transition to positive trust is of second
order, and the critical points are also
$\gamma=1/\left<k\right>$. Additionally, the values of average trust are
smaller than in networks with Poisson degree distribution and the same
average degree, for intermediary values of $\gamma$ after the
transitions. This is due to the smaller path multiplicity of graphs with
scale-free distribution: Even though the average shortest path length is
smaller in such graphs, the number of alternative paths is also smaller,
due to the dominance of vertices with smaller degree. Thus, if the
shortest path happens to have a small trust value, there will be a
higher probability there will not be an alternative path. In
Fig.~\ref{fig:perc} it is shown also the average best trust for $1 <
\tau < 2$, for which the average degree diverges. For such dense
networks, the values of $\left<s\right>$ are above zero for all values
of $\gamma > 0$, which is simply due to the fact that the average
shortest path length does not diverge in this case.

\section{The Pretty Good Privacy (PGP) Network}\label{sec:pgp}

In this section we investigate trust propagation on the Pretty Good
Privacy (PGP) network. In a broad manner PGP (or more precisely the
OpenPGP standard~\cite{thayer_openpgp_2007}) refers to a family of
computer programs for encryption and decryption of files, as well as
data authentication, i.e. generation and verification of digital
signatures.  It is often used to sign, encrypt and decrypt email. It
implements a scheme of public-key
cryptography~\cite{menezes_handbook_1996}, where the keys used for
encryption/decryption are split in two parts, one private and one
public. Both parts are related in way, such that the private key is used
exclusively for decryption and creation of signatures, and the public
key only for encryption and signature verification. Thus any user is
capable of sending encrypted messages and verifying the signature of a
specific user with her public key, but only this user can decrypt these
messages and generate signatures, using her private key, which she
should never disclose. The public keys are usually published in
so-called key servers, which mutually synchronize their databases, and
thus become global non-centralized repositories of public keys. However,
the mere existence of public key in a key server, associated with a
given identity (usually a name and an email address) is no guarantee
that this key really belongs to the respective person, since there is no
inherent verification in the submission process. This problem is solved
by the implementation of the so-called \emph{web of trust} of PGP keys,
whereby a user can attach a signature to the public key of another user,
indicating she trusts that this key belongs to its alleged owner. The
validity of a given key can then be inferred by transitivity, in a
self-organized manner, without the required presence of a central trust
authority. As such, this system represents an almost perfect example of
a trust propagation through transitivity.

As a rule, key signatures should only be made after careful
verification, which usually requires the two parties to physically meet.
Such a requirement transforms the web of trust into a snapshot of a
global social network of acquaintances, since the vast majority of keys
correspond to human users, which tend to sign keys of people with which
they normally interact. There is also a tendency to sign keys (upon
verification) from people which do not belong to a close circle of
acquaintances, with the sole purpose of strengthening the web of trust
with more connections. This tendency is well reflected by the so-called
``key signing parties'', where participants meet (usually after a large
technological conference) to massively sign each other's
keys~\cite{brennen_keysigning_2008}. Thus the structure of the PGP
network reflects the global dynamics of self-organization of human
peers in a social context.

This section is divided in two parts. In the first part we present some
aspects of the topology and temporal organization of the network. In the
second part we analyze the trust transitivity in the network, in view of
the trust metric we discussed previously.

\subsection{Network topology}

The PGP network used in this work was obtained from a snapshot of the
globally synchronized SKS key servers~\footnote{Available at {\tt
    http://key-server.de/dump/}} in November 2009. It is composed of
$N\approx2.5\times10^6$ keys and $E\approx7\times10^5$ signatures with a
very low average degree of $\left<j\right> = 0.28$. This means that many
keys are isolated and contain no signatures. Therefore we will
concentrate on the largest \emph{strongly connected component}, i.e. a
maximal set of vertices for which there is a path between any pair of
vertices in the set. The number of vertices $N\approx 4\times10^4$ in
this component is much smaller, but the network is much denser, with on
average $\left<j\right> \approx 7.58$ signatures per key (see summarized
data in table~\ref{tab:pgp-stats}). It represents the \emph{de facto}
web of trust, since the rest of the network is so sparsely connected
that no trust transitivity can be inferred from it. We note that keys
may have multiple ``subkeys'' which correspond to different identities
(usually different email addresses from the same person) and which can
individually sign other subkeys. For simplicity, in this work we have
collapsed subkeys into single keys, and possible multiple signatures
into a single signature. We have also discarded invalid, and revoked
keys and signatures.

\begin{table}[htb!]
  \begin{tabular}{llllll}
  $N$     & $E$    &$\left<j\right>$ & $r$     & $a$  & $c$  \\ \hline\hline
  2513677 & 703142 & $\approx 0.28$  & $0.45$  & $-0.02152(12)$ & 0.02321(9)\\
  39796   & 301498 & $\approx 7.58$  & $0.69$  & $0.0332(3)$ & $0.461(2)$\\ \hline
  \end{tabular}
  \caption{Summary of statistics for the whole PGP network (above) and the
    largest strongly connected component (below). $N$ is the number of vertices (keys),
    and $E$ is the number of edges (signatures), $\left<j\right>$ is the average
    in-degree, $r$ is the average reciprocity, $a$ is the
    assortativity coefficient and $c$ is the average clustering coefficient.
    \label{tab:pgp-stats}}
\end{table}

\begin{figure}[htb!]
  \includegraphics*{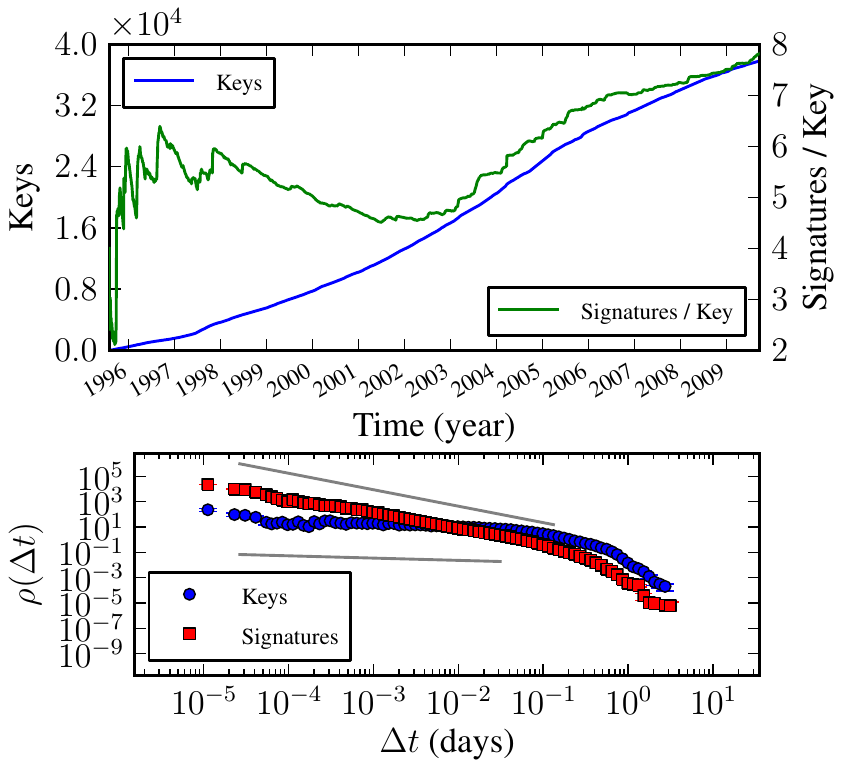}
  \caption{Number of keys and signatures as a function of time for the strongly
    connected component of the PGP network, and waiting time distribution
    between new keys and signatures. The straight lines are power-laws $\Delta t
    ^{-\xi}$, with $\xi=1.3$ (top) and $\xi=0.18$ (bottom).\label{fig:growth}}
\end{figure}

The number of keys and signatures in the strongly connected component
has been increasing over time, as shown in Fig.~\ref{fig:growth}. The
number of keys (which are now valid) was approximately the same for some
time and then slightly decreased for a period up to around 2002, and has
been increasing with an approximately constant rate since then. We note
that the number of keys may decrease since keys can expire or be
revoked. The number of signatures, on the other hand, seems to be
increasing with an accelerated rate, which is approximately constant,
and similar to the rate of growth of the number of keys. This means that
the average degree of the network is increasing with time, as can be
seen in Fig.~\ref{fig:growth}. Keys and signatures grow in an organized
manner, as shown by the waiting time distribution between the creation
of two subsequent keys or signatures, as shown in
Fig.~\ref{fig:growth}. These distributions are broad for several orders
of magnitude, from the order of seconds to days, approximately following
a power-law in this region. The fact that keys and signatures are often
created only seconds apart, and the waiting time distribution lacks any
discernible characteristic scale, except for a cut-off at large times
($\sim 1$ day), shows that the network does not grow in a purely random
fashion (which would generate exponentially-distributed waiting times),
and serves as a signature of an underlying organized growth process.

We will characterize the topology of the network by its degree
distribution and nearest-neighbours degree correlations, as well as
other standard network measures such as
clustering~\cite{newman_structure_2003},
reciprocity~\cite{zamora-lopez_reciprocity_2008} and community
structure~\cite{newman_finding_2004}. We will pay special attention to
the most highly connected vertices, some of which correspond to
so-called \emph{certificate authorities} and display a distinct
connectivity pattern, which has a special meaning for trust propagation.

\begin{figure}[htb!]
  \includegraphics*{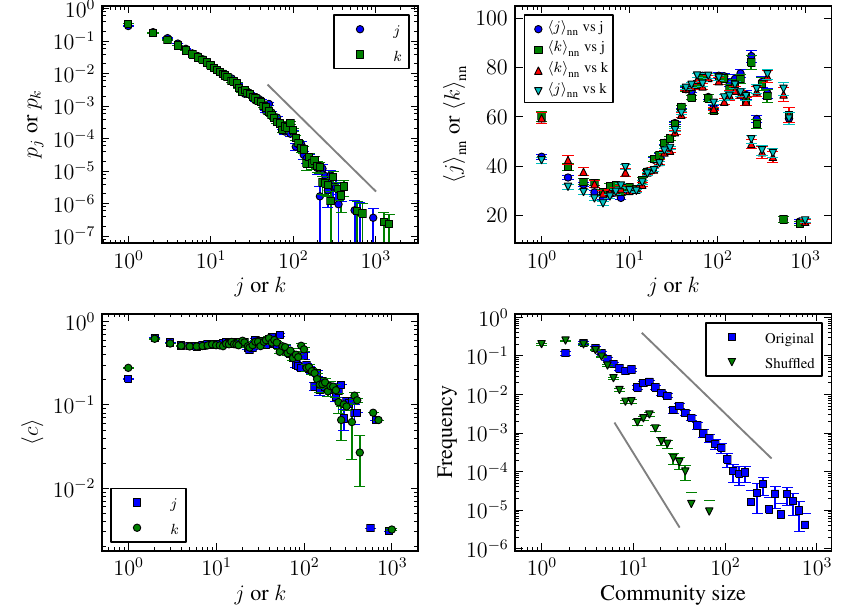}
  \caption{\label{fig:deg-stat} Several statistical properties for the
    PGP Network. {\bf Top left:} In- and out-degree distributions, $p_j$
    and $p_k$ respectively. The solid line corresponds to a power-law
    with exponent $2.5$.  {\bf Top right:} Average in- and out-degree of
    the nearest out-neighbours, as a function of the in- and
    out-degree. {\bf Bottom left:} Average lustering coefficient as a
    function of in- and out-degree. {\bf Bottom right:} Distribution of
    community sizes, for the unmodified and shuffled versions of the
    network. The solid lines correspond to power-laws with exponent
    $2.3$ (top) and $3.8$ (bottom).}
\end{figure}

\begin{figure}[htb!]
  \includegraphics*{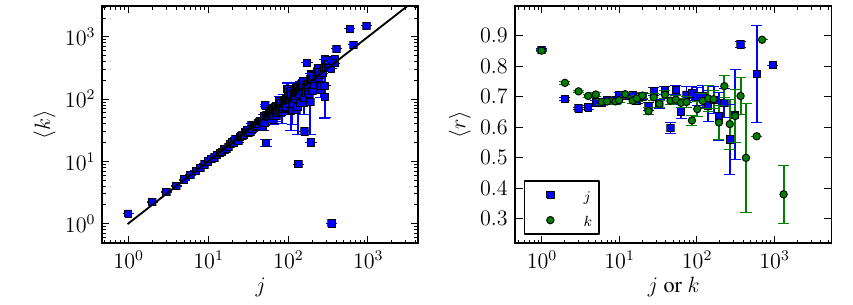}
  \caption{{\bf Left:} Average out-degree as a function of the in-degree of the
    same vertex. {\bf Right:} Average edge reciprocity, as a function of the in
    or out-degree of the source vertex. \label{fig:reciprocity}}
\end{figure}

\begin{figure}[htb!]
  \includegraphics*{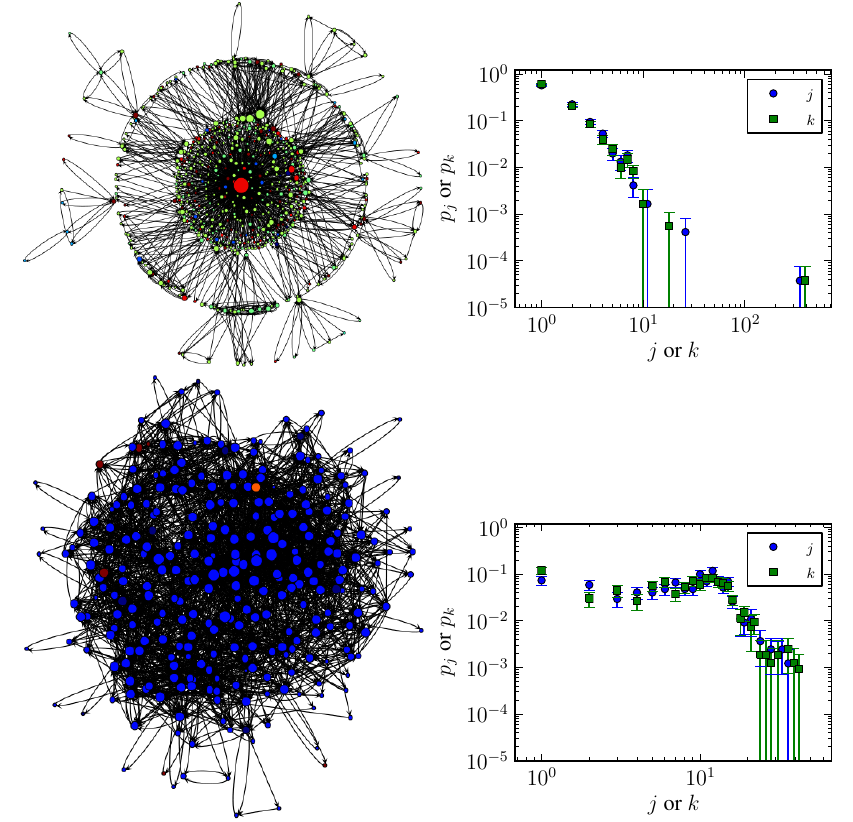}
  \caption{Two example communities of the PGP network, and their in- and
    out-degree distributions. The colors on the vertices correspond to
    the top-level domain (TLD) of the email addresses. {\bf Top:}
    Community containing the CACert.org certificate authority.  {\bf
      Bottom:} Community composed mostly of Austrian email addresses
    (.at TLD).\label{fig:community}}
\end{figure}

The network has very heterogeneous degree distributions, as can be seen
in Fig.~\ref{fig:deg-stat}, with some keys having on the order of $10^3$
signatures. They are possibly compatible with a power-law with exponent
$\sim 2.5$ for large degrees, but the distributions are not broad enough
for a precise identification.  The number of signatures on a given key
(the in-degree) and the number of signatures made by a the same key (the
out-degree) are strongly correlated, as can be seen in
Fig.~\ref{fig:reciprocity}, which shows the average out-degree $\avg{k}$
as a function of the in-degree $j$.
This is explained by the high reciprocity of the edges in the network,
i.e. if a key $a$ signs a key $b$, there is a very high probability that
key $b$ signs key $a$ as well. This is easy to understand, since key
verification usually requires physical presence, and both parties take
the opportunity to mutually verify each other keys in the same
encounter. The edge reciprocity~\cite{zamora-lopez_reciprocity_2008} is
quantified as the fraction $r = n^\leftrightarrow_e / E$, where
$n^\leftrightarrow_e$ is the number of reciprocal edges and $E$ is the
total number of edges in the network. The PGP network has a high value
of $r = 0.69$. The reciprocity is distributed in a slightly
heterogeneous fashion across the network, as is shown in
Fig.~\ref{fig:reciprocity}, where is plotted the average reciprocity of
the edges as a function of the in- and out-degrees of the source
vertex. It can be seen that the keys with very few signatures tend to
act in a very reciprocal manner, whereas the more prolific signers
receive less signatures back. This heterogeneity is further amplified
when one considers the degree correlation between nearest-neighbours, as
shown in Fig.~\ref{fig:deg-stat}, where it is plotted the average in-
and out-degree, $\avg{j}_\text{nn}$ and $\avg{k}_\text{nn}$, of the
nearest out-neighbours of the vertices in the network, as a function of
the in- and out-degree of the source vertex, $j$ and $k$.
The degree correlation shows an \emph{assortative} regime for
intermediary degree values ($\sim10$ -- $40$), meaning that vertices
with higher degrees are connected preferentially with other vertices
with high degree, but also some \emph{dissortative} features for
vertices with very high and very low degrees, where vertices with low
degree are connected preferentially with vertices with high degree, and
\emph{vice versa}. This mixed connectivity pattern leads to a very low
scalar assortativity coefficient~\footnote{%
  The scalar assortativity coefficient~\cite{newman_mixing_2003} is
  defined for an undirected graph as $a = \frac{1}{\sigma_q^2}\sum_{ij}
  ij\left(e_{ij} - q_{i} q_{j}\right)$ where $e_{ij}$ is the fraction of
  edges that connect vertices of degrees $i$ and $j$, $q_i =
  \sum_je_{ji}$ and $\sigma_q$ is the standard deviation of the
  distribution $q_i$. This definition yields values in the range
  $a\in[-1,1]$, with $a=-1$ for networks which are maximally
  dissortative, and $a=1$ for maximally assortative. For the PGP
  network, the direction of the edges was ignored in the calculation of
  $a$.}  of $a = 0.0332(3)$, which is unusual for social
networks~\cite{newman_why_2003}.  These differences become more clear
when one investigates more closely the keys with the largest degree in
the network, as it is shown in table~\ref{tab:hubs}.
\begin{table*}[htb!]
  \begin{tabular}{llllllll}
    \smaller
    Key ID &       Name  & $j$ & $k$ & $\left<j\right>_{\text{out}}$ & $c$ & Date \\ \hline\hline
    \texttt{D2BB0D0165D0FD58} & CA Cert Signing Authority (Root CA) <gpg@cacert.org>  & $ 965 $ & $ 1507 $ & $ 17.5(8) $ & $ 0.0031 $ & 2003-07-11 \\
    \texttt{2F951508AAE6022E} & Karlheinz Geyer (TUD) <geyerk.fv.tu@nds.tu-darmstadt.de>  & $ 661 $ & $ 744 $ & $ 59(2) $ & $ 0.0660 $ & 2004-12-07 \\
    \texttt{DBD245FCB3B2A12C} & ct magazine CERTIFICATE <pgpCA@ct.heise.de>  & $ 597 $ & $ 1348 $ & $ 18.3(12) $ & $ 0.0033 $ & 1999-05-11 \\
    \texttt{69D2A61DE263FCD4} & Kurt Gramlich <kurt@skolelinux.de>  & $ 406 $ & $ 644 $ & $ 71(3)) $ & $ 0.0807 $ & 2002-10-17 \\
    \texttt{948FD6A0E10F502E} & Marcus Frings <protagonist@gmx.net>  & $ 387 $ & $ 381 $ & $ 82(5)$ & $ 0.1110 $ & 2002-03-22 \\
    \texttt{29BE5D2268FD549F} & Martin Michlmayr <tbm@cyrius.com>  & $ 385 $ & $ 436 $ & $ 56(4)$ & $ 0.0499 $ & 1999-08-04 \\
    \texttt{566D362CEE0977E8} & Jens Kubieziel <jens@kubieziel.de>  & $ 369 $ & $ 414 $ & $ 73(4) $ & $ 0.1098 $ & 2002-08-23 \\
    \texttt{3F101691D98502C5} & Elmar Hoffmann <elho@elho.net>  & $ 352 $ & $ 1 $ & $ 348 $ & $ 0.1122 $ & 2005-02-17 \\
    \texttt{957952D7CF3401A9} & Elmar Hoffmann <elho@elho.net>  & $ 348 $ & $ 311 $ & $ 84(5) $ & $ 0.1086 $ & 2005-02-17 \\
    \texttt{CE8A79D798016DC7} & Josef Spillner <josef@coolprojects.org>  & $ 344 $ & $ 429 $ & $ 71(4) $ & $ 0.1007 $ & 2001-05-22 \\
    \texttt{89CD4B21607559E6} & Benjamin Hill (Mako) <mako@atdot.cc>  & $ 325 $ & $ 319 $ & $ 70(5) $ & $ 0.0801 $ & 2000-07-13 \\ \hline
  \end{tabular}
  \caption{\label{tab:hubs} The eleven keys with the largest number of
    signatures in the network, their respective in-degree $i$, out-degree $j$,
    average in-degree of the nearest out-neighbours $\left<j\right>_{\text{out}}$,
    clustering coefficient $c$, and date of creation.}
\end{table*}
As with the rest of the network, most of the largest keys belong to
individuals, with the exception of the first and third keys with the
most signatures, which belong to entities. These entities are known as
\emph{certificate authorities} and are created by organizations with the
intent of centralizing certification.  The largest authority is the
community-driven CAcert.org which issues digital certificates of various
kinds to the public, free of charge~\footnote{See the CAcert.org
  website: \texttt{http://cacert.org}}. The second largest authority is
the German magazine c't, which initiated a PGP certification campaign in
1997~\footnote{%
  A second, older c't key is also still among the largest hubs, with 289
  signatures. See \texttt{http://www.heise.de/security/dienste/
    Krypto-Kampagne-2111.html} for more details.} %
. These authorities interact with individuals in a different manner,
acting as a central mediator between loosely connected peers. This is
evident by the low clustering coefficient ($c \approx 0.003$), which is
one order of magnitude lower than the other (human) hubs ($c \sim 0.05$
-- $0.11$), and the average in-degree of their out-neighbours, which is
also significantly smaller than their human counterparts ($\sim 17$
vs. $60$ -- $80$, respectively). These different patterns represent
distinct paradigms of trust organization: Authority vs. Community-based;
each with its set of advantages and disadvantages. An authority-based
scenario relies on few universally trusted vertices which mediate all
trust propagation. In this way, the responsibility of key verification
is concentrated heavily on these vertices, which reduces the total
amount of verification necessary, and is thus more efficient. The most
obvious disadvantage is that the authorities represent central points of
failure: if an authority itself is not trusted, neither will be the keys
it certifies. Additionally, this approach may increase the probability
of forgery, since only one party needs to be deceived in order for
global trust to be achieved. The complementary scenario is the
community-based approach, where densely-connected clusters of vertices
provide certification for each other. This obviously requires more
diligence from the participants, but has the advantage of larger
resilience against errors, since the multiplicity of different paths
between vertices is much larger. In the PGP network both these paradigms
seem to be present simultaneously, as can be observed in detail by
extracting its community structure~\cite{newman_finding_2004}. This is
done by obtaining the community partition of the network which maximizes
the \emph{modularity} $Q$ of the network, defined as
\begin{equation}
  Q = \frac{1}{2E} \sum_{i\ne j} \left[A_{ij} - \frac{k_ik_j}{2E}\right]\delta(s_i, s_j),
\end{equation}
where $E$ is the total number of edges, $A_{ij}$ is the adjacency matrix
of the network, $k_i$ is the degree of vertex $i$, $s_i$ is the
community label of vertex $i$ and $\delta$ is the Kronecker
delta. According to this definition, a partition with high values of $Q$
is possible for networks with densely-connected groups of vertices, with
fewer connections between different groups. The maximum value of $Q=1$
is achieved only for "perfect" partitions of extremely segregated
communities. We note that the above definition is meaningful only for
\emph{undirected} graphs, and thus we apply it to the undirected version
of PGP network, where the direction of the edges is ignored. We used the
method of Reichardt et al~\cite{reichardt_statistical_2006} to obtain
the best partition, which resulted in modularity value of
$Q\approx0.73$. As a comparison, we computed the modularity for a
shuffled version of the network, where the edges were randomly placed,
but the degrees of the vertices were preserved, which resulted in the
significantly smaller value $Q\approx0.03$. The distribution of
community sizes seems to have a power-law tail with exponent $\sim 2.3$
($\sim 3.8$ for the shuffled network), characterizing a scale-free
structure. By isolating the individual communities, one can clearly see
strong differences between those in the vicinity of the certificate
authorities and ``regular'' communities. In Fig.~\ref{fig:community} is
shown two representative examples of these two types of communities: On
top is the community around the CAcert.org certificate authority, and is
composed of $677$ keys, with an average $6.9$ signatures per key. Its
degree distributions are shown on the side, from which the large
discrepancy between the most central vertex and the rest of the
community can be observed. The colors on the vertices correspond to the
Top-Level Domain (TLD) of the email addresses associated with each key,
and serve as an indication of the geographical proximity of the
individuals. For the community containing CAcert.org, a high degree of
geographical heterogeneity is present. This is corroborated also by the
fact that there are fewer direct edges between individuals. On the
bottom of Fig.~\ref{fig:community} it is shown a community composed
almost exclusively of keys with Austrian email addresses (.at TLD) which
show a completely different pattern, lacking any central authority. It
is smaller, with $287$ keys, but denser, with $10$ signatures per
key. This pattern is repeated for most of the largest communities in the
graph. Some non-centralized communities have a broader degree
distribution than the Austrian community, but only those associated with
certificate authorities display a centralized pattern such as in the top
of Fig.~\ref{fig:community}.

We now turn to the trust propagation on the PGP network.

\subsection{Trust transitivity}

In order to properly investigate trust transitivity in the PGP network,
it is necessary to know the direct trust values associated with each
signature, which indicate the level of scrutiny in the key verification
process. The OpenPGP standard~\cite{thayer_openpgp_2007} defines four
trust ``classes'' for signatures, according to the degree of
verification made. Unfortunately, these classes are universally ignored,
and most signatures fall into the ``generic'' class, from which no
assertion can be made. Since the actual level of verification of the
keys is in fact unknown, we will investigate hypothetical situations
which represent different strategies the PGP users may use to verify
keys. In the last section we have shown that the network is composed of
different connection patterns: community clusters and centralized trust
authorities. Depending on how these connection patterns are judged more
trustworthy, the values of transitive trust will be different. Here we
will consider three possible scenarios: 1. Random distribution,
2. Authority-centered trust, and 3. Community-centered trust. In all
situations we will consider that all signatures have the same trust
value of $c=1/2$, except for a fraction $\gamma$ of edges which have
absolute trust $c=1$, which is selected as follows for each situation:
\begin{enumerate}
\item \emph{Random:} The $\gamma E$ edges are chosen randomly among all $E$
  edges.
\item \emph{Authority-centered:} The $\gamma E$ edges with the largest

  \emph{betweenness}~\cite{linton_c._freeman_set_1977} $b_e$ are chosen,
  which is defined as
  \begin{equation}
    b_e = \sum_{i\ne j} \frac{\sigma_{ij}(e)}{\sigma_{ij}},
  \end{equation}
  where $\sigma_{i,j}$ is the number of shortest paths from vertex $i$ to $j$,
  and $\sigma_{ij}(e)$ is the number of these paths which contain the edge
  $e$. This distribution favours edges adjacent to nodes with high degree, and
  also edges which bridge different communities.
\item \emph{Community-centered:} The $\gamma E$ edges with the largest
  \emph{edge clustering} $\tau_e$ are chosen, which is defined as
  \begin{equation}
    \tau_e = \frac{\sum_{i}A_{s(e),i}A_{i,t(e)}}{\sqrt{k_{s(e)}j_{t(e)}}},
  \end{equation}
  where $s(e)$ and $t(e)$ are the source and target vertices of edge
  $e$, $A_{i,j}$ is the adjacency matrix, and $j_i$ and $k_i$ are the
  in- and out-degrees of vertex $i$, respectively. This quantity
  measures the density of out-neighbours of the $s(e)$ which are also
  in-neighbours of $t(e)$, and simultaneously the density of
  in-neighbours of $t(e)$ which are out-neighbours of $s(e)$. This
  distribution favours edges with belong to densely-connected
  communities. For instance, the edges of a \emph{clique} (i.e. a
  complete subgraph) will all have the value $\tau_e= 1-1/(n-1)$, where
  $n$ is the size of the clique, which will approach the maximum value
  $\tau_e\to 1$ for a sufficiently large clique size.
\end{enumerate}

\begin{figure}[htb!]
  \includegraphics*{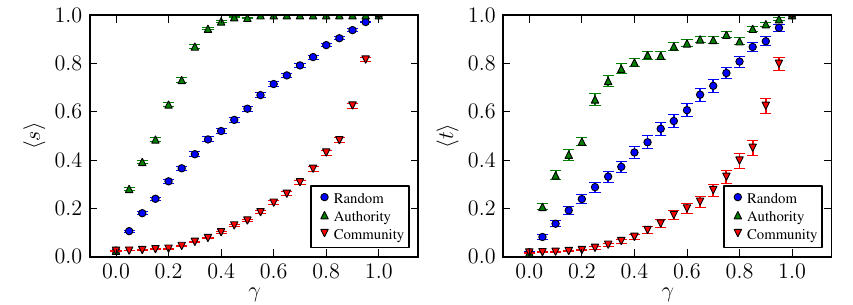}
  \caption{Average best trust $\avg{s}$ and pervasive trust $\avg{t}$, as a
    function of the fraction of edges with absolute trust $\gamma$, for the PGP
    network. The different curves correspond to the different trust distribution
    scenarios described in the text.\label{fig:pgp-avg-t}}
\end{figure}
\begin{figure*}[hbt!]
  \includegraphics*{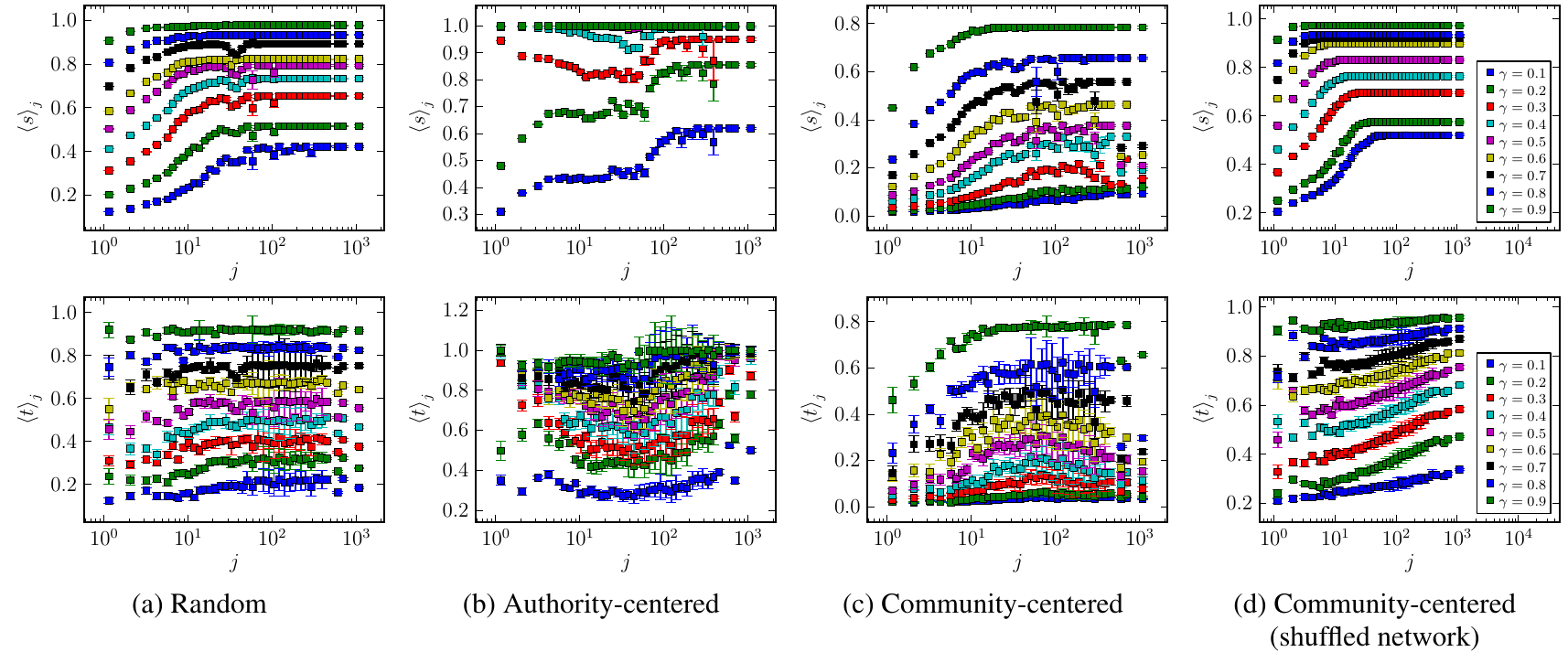}
  \caption{\label{fig:t-corr} Average best trust $\avg{s}$ and pervasive
    trust $\avg{t}$, as a function of the in-degree $j$ and the fraction
    of edges with absolute trust $\gamma$, for the PGP network. The
    different plots correspond to the different trust distribution
    scenarios described in the text: (a) Random distribution, (b)
    authority-centered distribution and (c) community-centered
    distribution. The plots (d) correspond to a community-centered
    distribution, done on a shuffled version of network, with the same
    degree sequence.}
\end{figure*}

In Fig.~\ref{fig:pgp-avg-t} it is shown the average best trust
transitivity, Eq.~\ref{eq:lotta} and average pervasive trust
Eq.~\ref{eq:pervasive} for the PGP network, as a function of $\gamma$
according to the different approaches. We note that, due to the
relatively small size of the network, no discontinuous transition is
seen.  The authority-centered trust leads to significantly higher values
of $\avg{s}$ and $\avg{t}$, and the community-based distribution to the
lowest values. This is expected, since distributing trust according to
the edge betweenness essentially \emph{optimizes} trust transitivity,
putting the highest values along the shortest paths between
vertices. The community-centered approach does exactly the opposite,
favoring intra-community connections, and results in the lowest values
of average trust. Thus, favoring the hubs and authorities is clearly
more \emph{efficient}, if the objective is solely to increase the
average trust in the network. However, pure efficiency may not be what
is desired, since it relies in the opinion of a much smaller set of
vertices, which eases the job of dishonest parties, which need only to
convince these vertices in order to be trusted by a large portion of the
network. Some of these issues become more clear by observing how nodes
with different degrees receive trust with each of these strategies, as
show in Fig.~\ref{fig:t-corr}.  For random distribution of trust, the
vertices with higher degree receive a natural bias in the values of
average best in-trust, $\avg{s}$, since the shortest paths leading to
them tend to be smaller. But the fair nature of the definition of $t$
compensates for this, and the values of $\avg{t}$ are almost independent
of the in-degree of the vertices. The highly connected nodes become more
trusted only with the authority-centred approach. Interestingly, in this
situation the nodes with the \emph{smallest} degrees also receive a
large value of trust, since most of them are ``fringe'' nodes connected
only with the hubs (see Fig.~\ref{fig:deg-stat}).  The vertices with
intermediary degrees are thus left in the limbo, and are in effect
\emph{penalized} for their community pattern. The almost symmetrically
opposite situation is obtained with the community-centered trust
distribution, where both the vertices with smallest and largest degrees
receive the smallest trust values, and the intermediary nodes are judged
more trustworthy due to their strong communities. We note that this
effect is not due simply to the way the values of trust are distributed,
but depend strongly on the existence of communities in the network. This
is evident when the same trust distribution is applied to a shuffled
version of the network, with the same degree sequence, as is shown in
Fig.~\ref{fig:t-corr}. For such a network, the community structure
disappears, and the highly connected nodes come again in the lead.

\section{Conclusion}\label{sec:conclusion}

We investigated properties of trust propagation on network based on the
notion of trust transitivity. We defined a trust metric, called
\emph{pervasive trust} which provides inferred trust values for pairs of
nodes, based on a network of direct trust values. The metric extends
trust transitivity to the situation where multiple paths between source
and target exist, by combining the best trust transitivity to the
in-neighbours of a given target node, and their direct trust to the
target. The trust values so-obtained are unbiased, personalized and well
defined for any possible network topology. Equipped with this metric we
analyzed the conditions necessary for global trust propagation in large
systems, using random networks with arbitrary degree distributions as a
simple model. We analytically obtained the average best trust
transitivity (as well as pervasive trust) as a function of the fraction
$\gamma$ of edges with \emph{absolute} trust $c=1$. We found that there
is a specific value of $\gamma = \gamma^*$, below which the average
trust is always zero. For $\gamma \geq \gamma^*$ the average value jumps
discontinuously to a positive value.

Using the defined trust metric, we investigated trust propagation in the
Pretty Good Privacy (PGP) network~\cite{guardiola_macro-_2002,
  boguna_models_2004}. We gave an overview of the most important
topological and dynamical features of the PGP network, and identified
mixed connectivity patters which are relevant for trust propagation:
namely the existence of trust authorities and of densely-connected
non-centralized communities. Based on these distinct patterns, we
formulated different scenarios of direct trust distribution, and
compared the average inferred trust which results from them. We found
that an authority-centered approach, where direct trust is given
preferentially to nodes which are more central, leads to a much larger
average trust, but at the same time benefits nodes at the fringe of the
network, which are only connected to the authority hubs, and for which
no other information is available. Symmetrically, a community-centered
approach, where edges belonging to densely-connected communities are
favoured with more trust, results in less overall trust, but both the
fringe nodes and the authorities receive significantly less trust than
average. These differences are not simply due to the different ways the
direct is distributed, but rather to the fact that the dense communities
and the trust authorities are somewhat segregated. These differences
illustrate the advantages and disadvantages of both paradigms of trust
propagation, which seem to be coexist in the PGP network. It also serves
as an insightful example of how dramatically the direct trust
distribution can influence the inferred trust, even when the underlying
topology remains the same.

In this work, we have concentrated on static properties of trust
propagation. However most trust-based systems are dynamic, and change
according to some rules which are influenced by the trust propagation
itself. One particularly good example is market
dynamics~\cite{nicholaas_j._vriend_self-orgainzed_1994,
  anand_financial_2009, bornholdt_handbook_2003} where sellers (or
borrowers) do not perform well if they have a poor track record, which
will be partially influenced by trust. Thus, it remains to be seen how
trust transitivity can be carried over to such types of models, and what
role it plays in shaping their dynamics.

\section*{Acknowledgments}
We thank Alexandre Hannud Abdo for carefully reading the
manuscript. This work has been supported by the DFG under Contract
No. Dr300/5-1.

\bibliography{clean_bib}

\end{document}